\documentclass[aps,prl,twocolumn,superscriptaddress,floatfix,nofootinbib]{revtex4-2}
\usepackage[T1]{fontenc}
\usepackage{amsmath,amssymb,bm,braket}
\usepackage{graphicx}
\usepackage{tikz}
\usetikzlibrary{arrows.meta,calc,decorations.pathmorphing,positioning,shapes.geometric}
\usepackage[colorlinks=true,allcolors=blue]{hyperref}
\newcommand{\pp}{\mathbf p}
\newcommand{\rr}{\mathbf r}

\begin{document}

\title{Exact Electron Dynamics in Linear and Chiral Cavities:\\
Complete Field versus Single-Helicity Reduction}
\author{Patrick Barron}
\author{Kalman Varga}
\email{kalman.varga@vanderbilt.edu (corresponding author)}
\affiliation{Department of Physics and Astronomy, Vanderbilt University, Nashville, Tennessee 37235, United States}

\begin{abstract}
Can a chiral cavity turn a free electron into a Landau-like orbiting particle?  We answer with exact solutions of three quadratic electron--cavity models in the dipole limit, valid at arbitrary coupling.  A single linearly polarized mode dresses the electron mass, which delays its flight, slows the spreading of a wavepacket and entangles the packet with the cavity.  The complete transverse field, with both circular helicities, keeps two commuting, gapless translations at any coupling, mode number and helicity splitting, so the electron stays ballistic and no synthetic magnetic field appears.  Keeping a single helicity gives two inequivalent models that are not related by a gauge transformation: reducing before the Power--Zienau--Woolley transformation gives ballistic motion, whereas reducing after it gives a Landau-like crossed-quadrature Hamiltonian.  They differ by a real uniform field $B=s\lambda^2/(2\omega_c)$ ($\omega_c$ the mode frequency, $\lambda$ the coupling, $s=\pm1$ the helicity).  In the complete theory the frequency $\omega_{-s}$ of the partner helicity, measured against $g^2=\lambda^2/(2\omega_c)$, decides between them: the electron is ballistic for a stiff partner, and Landau-like only at the soft-partner point, through an exact canonical reorganization that holds for the active sector but not for observables such as the electron--photon entanglement.  Projecting onto one helicity after the PZW transformation is therefore not by itself a mechanism for synthetic magnetism; the mode content of the cavity is.
\end{abstract}
\maketitle

Confining the electromagnetic field so tightly that its vacuum fluctuations reshape matter is an established route to material control without external driving~\cite{garciavidal2021,schlawin2022,bloch2022}.  Such effects are largest in the ultrastrong-coupling regime, where the light--matter coupling becomes comparable to the mode frequency and perturbative treatments fail; this regime has been reached experimentally, notably in circuit quantum electrodynamics and in cavity-coupled two-dimensional electron gases, where magnetotransport is controlled by Landau polaritons~\cite{Scalari2012,ParaviciniBagliani2019} and the topological protection of the integer quantum Hall effect breaks down~\cite{Appugliese2022}.  Free electrons occupy a distinctive place in this landscape: an extended wavepacket with a continuous spectrum can exchange arbitrary multiples of the photon energy with a confined field, which underlies free-electron quantum optics~\cite{GarciaDeAbajo2010,Ruimy2025FreeElectronQuantumOptics} and its demonstrations of coherent electron--photon interaction~\cite{barwick2009,feist2015,wang2020,kfir2020,Dahan2021,Jung2023,Kfir2019Entanglement}.  The elementary question is what a free electron wavepacket does inside a cavity.

Exactly solvable reference problems are valuable in the ultrastrong-coupling regime, where most approaches are perturbative or numerical, but they are scarce~\cite{Rokaj2022FreeElectronGas}.  For chiral cavities, time-reversal breaking has been proposed as a way to engineer electronic properties with vacuum fields alone, and cavity fluctuations modify the magnetic response of quantum Hall states~\cite{Yang2026QuantumHallChiralCavity}.  A closed-form answer to how a single wavepacket drifts, spreads and entangles in linearly and circularly polarized modes, and whether a chiral vacuum can act on a free charge as a magnetic field at all, has nevertheless been missing.

In this Letter we show that a free electron in a cavity admits three analytically solvable models at arbitrary coupling in the dipole limit, which differ in the field content retained and in how the field is reduced; they are derived from minimal coupling below.  Case I, a single linearly polarized mode, has an exact operator motion with a dressed mass, a time-of-flight delay and a reduced spreading rate.  Case II, the complete transverse field with both circular helicities, keeps two commuting, gapless translations at any coupling, mode number and helicity splitting.  Case III reduces the field to one helicity and yields two inequivalent effective models, a ballistic one (III-V) and a Landau-like crossed-quadrature one (III-L); we derive their exact relation and the partner-helicity frequency that separates them.  Together the cases give a fully analytic map between complete-field, single-helicity and Landau-like physics, with experimentally distinct signatures for each.  Experimentally motivated chiral cavities~\cite{Voronin2022SingleHandedness,Paterson2025HelicityTwisted,Crew2026UPLOADHELIX} make the field content of a ``chiral cavity'' a physical question.

The common starting point is the minimal-coupling Hamiltonian
\begin{equation}
\hat H=\frac{1}{2m}\left[\hat{\pp}-q\hat{\mathbf A}(\hat{\rr})\right]^2+\hat H_{\rm cav}.
\label{eq:start}
\end{equation}
We set $\hbar=m=1$ below.  In the long-wavelength approximation, $\hat{\mathbf A}(\hat{\rr})\to\hat{\mathbf A}(\mathbf r_0)$, so a free-electron velocity-gauge Hamiltonian has no explicit electron-coordinate dependence.

The three models follow from Eq.~\eqref{eq:start} by the choice of cavity field, as sketched in Fig.~\ref{fig:threecases}: Case I, a single linearly polarized mode; Case II, the complete transverse field containing both circular helicities; and Case III, a deliberate reduction of the complete field to one helicity.  Case III has two versions that differ in the order of operations: projecting to one helicity before the PZW transformation gives the velocity-gauge model (III-V), and projecting after it gives the length-gauge model (III-L).  The vector potentials in the long-wavelength limit are
\begin{equation}
\begin{aligned}
\text{Case I:}\quad &\hat{\mathbf A}=A_0\hat Q\,\hat{\mathbf z}, && qA_0=\sqrt2\,g,\\
\text{Case II:}\quad &\hat{\mathbf A}=A_0\big(\hat Q_x\hat{\mathbf x}+\hat Q_y\hat{\mathbf y}\big), && qA_0=\sqrt2\,g,\\
\text{Case III:}\quad &\hat{\mathbf A}=A_0\big(\hat Q\,\hat{\mathbf x}-s\hat P\,\hat{\mathbf y}\big), && qA_0=g,
\end{aligned}
\label{eq:Acases}
\end{equation}
where $\hat Q=(\hat a+\hat a^\dagger)/\sqrt2$ and $\hat P=(\hat a^\dagger-\hat a)/(i\sqrt2)$ are the dimensionless quadratures of a cavity mode whose frequency is $\omega_c$ (the subscript stands for ``cavity''; it is not a cyclotron frequency), with $[\hat Q,\hat P]=i$, $g$ is the dimensionless electron--photon coupling strength, $\hat Q_x$ and $\hat Q_y$ belong to two independent modes, and $s=\pm1$ labels the retained helicity.  The components of $\hat{\mathbf A}$ commute in Cases I and II; in Case III they are conjugate quadratures of one oscillator, $[A_x,A_y]=-isA_0^2$.

\begin{figure*}[t]
\centering
\begin{tikzpicture}[x=1cm,y=1cm,>=Latex,font=\footnotesize,
  mirror/.style={fill=gray!45,draw=gray!60},
  panel/.style={rounded corners=3pt,fill=gray!5,draw=gray!45},
  pkt/.style={draw=red!60!black,fill=red!15},
  help/.style={draw=red!65!black,thick,-{Latex[length=2mm]}},
  helm/.style={draw=blue!65!black,thick,-{Latex[length=2mm]}}]
\def\H{4.7}\def\Wd{5.7}
\foreach \X/\ttl in {0/{\textbf{Case I}\ \ one linear mode},6.0/{\textbf{Case II}\ \ complete field, both helicities},12.0/{\textbf{Case III}\ \ one retained helicity}}{
  \draw[panel] (\X,0) rectangle ++(\Wd,\H);
  \node at (\X+\Wd/2,\H-0.30) {\ttl};
  \fill[mirror] (\X+0.25,1.55) rectangle ++(0.22,2.35);
  \fill[mirror] (\X+\Wd-0.47,1.55) rectangle ++(0.22,2.35);
}
\begin{scope}
 \foreach \x/\l in {1.2/0.45,2.1/0.85,3.0/0.85,3.9/0.45}{\draw[<->,blue!65!black,thick] (\x,2.72-\l/2) -- (\x,2.72+\l/2+0.2);}
 \node[blue!65!black] at (2.55,3.78) {$\mathbf E\parallel\hat z$};
 \draw[->,gray!70,dashed,thick] (1.9,1.65) -- (2.8,2.55);
 \node[gray!60!black,anchor=east] at (1.75,2.3) {free};
 \draw[->,red!65!black,thick] (1.9,1.65) -- (3.3,2.05);
 \node[red!65!black,anchor=west] at (3.35,2.0) {dressed};
 \draw[pkt] (1.9,1.65) ellipse (0.3 and 0.2);
 \node[align=center] at (2.85,0.95) {$v_z\to v_z/m^{*},\ v_\perp$ unchanged};
 \node[align=center] at (2.85,0.4) {$m^{*}=\mathrm{diag}(1,1,1{+}\lambda^2/\omega_c^2)$};
\end{scope}
\begin{scope}[shift={(6.0,0)}]
 \draw[helm] (1.15,3.25) arc[start angle=180,end angle=500,x radius=0.55,y radius=0.28];
 \draw[help] (3.35,3.25) arc[start angle=0,end angle=320,x radius=0.55,y radius=0.28];
 \node[red!65!black] at (3.35,3.78) {$\sigma^{+}$};
 \node[blue!65!black] at (1.15,3.78) {$\sigma^{-}$};
 \node at (2.25,3.25) {$+$};
 \draw[->,black,thick] (1.7,1.9) -- (3.9,1.9);
 \draw[pkt] (2.2,1.9) ellipse (0.3 and 0.2);
 \node[align=center] at (2.85,1.2) {$[A_x,A_y]=0,\ \ [C_x,C_y]=0$};
 \node[align=center] at (2.85,0.62) {two gapless translations, isotropic $m^{*}$};
 \node[align=center] at (2.85,0.22) {no Landau gap};
\end{scope}
\begin{scope}[shift={(12.0,0)}]
 \draw[help] (2.85,3.52) arc[start angle=270,end angle=-50,x radius=0.55,y radius=0.28];
 \node[red!65!black,anchor=west] at (3.45,3.5) {one $\sigma^{s}$};
 \draw[gray!50] (2.85,2.15) -- (2.85,0.18);
 \node[align=center] at (1.6,2.0) {\textbf{III-V}\\[-1pt]{\scriptsize project, then PZW}};
 \draw[pkt] (0.8,1.2) ellipse (0.28 and 0.18);\draw[->,thick] (0.8,1.2) -- (2.45,1.2);
 \node at (1.6,0.62) {ballistic};
 \node[align=center] at (4.2,2.0) {\textbf{III-L}\\[-1pt]{\scriptsize PZW, then proj.}};
 \draw[->,thick,domain=0:330,smooth,variable=\a] plot ({4.2+0.55*cos(\a)},{1.25+0.38*sin(\a)});
 \node[align=center] at (4.2,0.42) {Landau-like,\\$B_{\rm syn}=sg^{2}$};
\end{scope}
\end{tikzpicture}
\caption{The three cases studied. \textbf{Case I}: a single linearly polarized mode couples one electron axis, and the cavity dresses only the velocity component along $\hat z$ by the factor $1/m^{*}$ (dashed: free motion; solid: dressed). \textbf{Case II}: the complete transverse field contains both circular helicities $\sigma^{\pm}$; their noncommuting quadratures cancel, $[A_x,A_y]=0$, so the electron keeps two commuting, gapless translations. \textbf{Case III}: retaining one helicity maps the two field components onto one oscillator. Projecting before the PZW transformation (III-V) gives ballistic motion; projecting after it (III-L) gives the Landau-like Hamiltonian with $B_{\rm syn}=sg^{2}$, which requires a real co-handed field, a softened partner helicity, or directly engineered couplings.}
\label{fig:threecases}
\end{figure*}

Case I: one linearly polarized mode.  For the potential of Eq.~\eqref{eq:Acases}, writing $\lambda^2=2g^2\omega_c$, direct substitution gives
\begin{equation}
\hat H_{\rm lin}^{V}=\frac12(\hat p_z-\sqrt2g\hat Q)^2+\frac{\omega_c}{2}(\hat Q^2+\hat P^2).
\label{eq:lin}
\end{equation}
The momentum $p_z$ is conserved.  With $\Omega=\sqrt{\omega_c^2+\lambda^2}$, the exact position operator has the form
\begin{equation}
\hat z(t)=\hat z(0)+\frac{\hat p_z}{m^*}t+\hat z_{\rm osc}(t),\qquad
m^*=1+\frac{\lambda^2}{\omega_c^2},
\label{eq:linmotion}
\end{equation}
where $z_{\rm osc}$ is bounded and periodic at $\Omega$ (explicit form: see SM Sec.~S2).  Hence $\mathrm{Var}\,z(t)\sim\mathrm{Var}(p_z)t^2/m^{*2}$.  Because the mass tensor is $\mathrm{diag}(1,1,m^*)$, the direction of the mean velocity of a three-dimensional beam, measured from the polarization axis, follows as a corollary,
\begin{equation}
\tan\theta_{\rm out}=m^*\tan\theta_{\rm in}.
\label{eq:deflection}
\end{equation}
Equations~\eqref{eq:linmotion} and~\eqref{eq:deflection} say that a single linearly polarized mode makes a free charge heavier along its polarization axis only (SM Fig.~1).  The dispersion-based mass of a charge coupled to cavity modes in the long-wavelength limit is known~\cite{Rokaj2022FreeElectronGas,Welakuh2023MassRenorm,FrohlichPizzo2010}; exact solutions of an electron in a single quantized plane-wave mode are older still~\cite{BergouVarro1981}; the exact Heisenberg solution here adds the full operator motion, including the bounded $z_{\rm osc}$.  The canonical momentum $p_z$ is conserved, yet the kinetic velocity along $z$ is $p_z/m^*$ because the charge drags the mode quadrature $Q$ with it, so the mean velocity inside the dressed region rotates away from the polarization axis.  The drift coefficient holds at arbitrary coupling and for any photon state; only $z_{\rm osc}$ depends on the state.  Two predictions hold however the beam enters and leaves the cavity, which the uniform dipole model does not describe.  First, crossing a dressed length $L$ along $z$ takes the time $m^*L/p_z$ instead of $L/p_z$, a delay of $(m^*-1)L/p_z$.  Second, the wavepacket spreads $m^*$ times more slowly.

Gaussian packet.  A packet is a superposition of conserved-$p_z$ components, each evolving under Eq.~\eqref{eq:lin} with $\hat p_z$ replaced by its eigenvalue $p_z$, a cavity oscillator driven by the linear force $\sqrt2\,g\,p_z$ on $Q$.  Its centre moves at $\bar p_z/m^*$, and for a minimum-uncertainty packet with the photon in vacuum and no initial position--momentum correlation, \begin{equation}\mathrm{Var}\,z(t)=\mathrm{Var}\,p_z\Big[\frac{t^2}{m^{*2}}+\frac{2\lambda^2}{m^*\Omega^3}\,t\sin\Omega t\Big]+R(t),\label{eq:varz}\end{equation} where $R(t)$ is bounded in time (it does not grow): it collects $\mathrm{Var}\,z(0)$, a term $\propto\mathrm{Var}\,p_z\sin^2\Omega t$, and the photon-fluctuation terms (see SM Sec.~S2.B).  The $t^2/m^{*2}$ law is therefore accompanied by one correction that grows linearly in $t$ with an oscillating coefficient, the $t\sin\Omega t$ term, and by the bounded remainder $R(t)$.  A packet also entangles with the cavity, which a plane wave never does: each $p_z$ component drives the mode to a different displaced state, so the reduced photon state is a mixture whose entropy grows with $\mathrm{Var}\,p_z$ and returns to zero whenever $\Omega t=2\pi n$.

Many modes.  Because the field uses one real quadrature, the PZW transformation is a pure shift whose Baker--Campbell--Hausdorff series stops at first order, so it is residual-free and Case~I is a gauge-unambiguous benchmark (SM Sec.~S2).  The solution extends to $N$ real-polarized modes with mass tensor $m^*=1+\sum_k(2g_k^2/\omega_k)\,\mathbf e_k\mathbf e_k^{T}$ (see SM Sec.~S2.A): the softest and most strongly coupled modes dominate, and because all $A_i$ commute for real polarizations, no number of such modes can produce a noncommuting algebra, which is the origin of the complete-field no-go result, Eq.~\eqref{eq:nogo} below.  For a continuum the sum is absorbed into the observable mass, so a physical cavity effect is the difference between the cavity and free-space mode sums.

{\it Significance and applications} (Case I).  Case I is first an exactly solvable benchmark: at arbitrary coupling and photon state it is an exactly known single-electron limit against which approximate first-principles treatments (mean-field Maxwell--Kohn--Sham, approximate photon functionals, free-space mass-renormalization estimates~\cite{Welakuh2023MassRenorm}) can be checked, in particular for the fluctuation and entanglement observables that a product-state treatment cannot capture; the quadratic terms whose omission makes the problem ill-defined~\cite{Rokaj2018NoGroundState,Schafer2020Quadratic} are retained.  For one mode of effective volume $V$ the shift is $m^*-1=2g^2/\omega_c=q^2/(\epsilon_0 m V\omega_c^2)$ in SI units, of order $10^{-12}$ for a single electron in a centimetre-scale microwave cavity and $10^{-6}$ for a nanophotonic mode of volume $10^{-3}\lambda^3$ (see SM Sec.~S2.C).  The delay and the reduced spreading are therefore detectable for free-electron beams only in deeply sub-wavelength or collectively enhanced geometries, and directly accessible in platforms with large effective coupling, such as circuit and semiconductor ultrastrong-coupling systems~\cite{FornDiaz2019USC,Kockum2019USC}.  Delays measured for several beam directions relative to the polarization axis reconstruct the full mass tensor, and oscillations of the electron--photon entropy at the dressed frequency provide a time-domain spectroscopy of the polaritonic modes (see SM Sec.~S10.A).

Case II: complete transverse field.  The second case keeps the complete transverse field.  The simplest balanced cavity uses two independent Cartesian oscillators with the potential of Eq.~\eqref{eq:Acases} and
\begin{equation}
\hat H_{\rm cav}=\frac{\omega_c}{2}\sum_{\mu=x,y}(Q_\mu^2+P_\mu^2),
\label{eq:completeA}
\end{equation}
with the same Cartesian coupling as in Eq.~\eqref{eq:lin}, $[Q_i,P_j]=i\delta_{ij}$ and therefore $[A_x,A_y]=0$.  The circular operators
\begin{equation}
a_\pm=\frac{a_x\mp ia_y}{\sqrt2}
\label{eq:circ}
\end{equation}
are only a change of basis: both helicities are retained.  The velocity-gauge Hamiltonian is
\begin{equation}
\begin{split}
\hat H_{2h}^{V}={}&\frac12(p_x-\sqrt2gQ_x)^2+\frac12(p_y-\sqrt2gQ_y)^2\\
&+\frac{\omega_c}{2}\sum_{\mu=x,y}(Q_\mu^2+P_\mu^2).
\end{split}
\label{eq:twoh}
\end{equation}
Immediately,
\begin{equation}
[p_x,H_{2h}^{V}]=[p_y,H_{2h}^{V}]=0,\qquad[p_x,p_y]=0.
\label{eq:nogo}
\end{equation}
Thus the complete transverse dipole field leaves a free charge with two commuting, gapless translation generators: this is the no-go result for synthetic magnetism from a complete field, since commuting translations cannot describe a uniform magnetic field.  Related gauge-invariance no-go theorems for photon condensation and superradiance~\cite{Nataf2010,Andolina2019} concern a different quantity.  This conclusion is independent of polarization basis and extends to any number of spatially uniform modes.

The same statement is transparent in length gauge.  Because $Q_x$ and $Q_y$ commute, the complete-field PZW unitary
\begin{equation}
U=\exp[i\sqrt2 g(xQ_x+yQ_y)]
\label{eq:U2h}
\end{equation}
is an ordinary canonical transformation, and the conserved translations become commuting hybrid generators $C_x=p_x+\sqrt2gQ_x$ and $C_y=p_y+\sqrt2gQ_y$, with $[C_x,C_y]=0$.  The electron remains ballistic with the isotropic $m^*$ of Eq.~\eqref{eq:linmotion}.

If the cavity distinguishes the two helicities, a gyrotropic term such as $\Delta(Q_xP_y-Q_yP_x)$ gives $\omega_\pm=\omega_0\pm\Delta$ and the exact dressed mass $m^*=1+g^2/\omega_++g^2/\omega_-$ (see SM Sec.~S3).  This changes the chiral response but not Eq.~\eqref{eq:nogo}: chirality can manifest only once translation invariance is broken, through anomalous velocity under a force, circular dichroism of a trapped charge, or cavity dressing in a real magnetic field.

{\it Significance and applications} (Case II).  Case II is the complete-field reference for chiral and time-reversal-breaking cavities~\cite{Yang2026QuantumHallChiralCavity,Hubener2021ChiralCavities,Andberger2024THzChiral,Kulkarni2025}: it fixes exactly what a spatially uniform cavity does to a free charge.  An electron gas in a gyrotropic cavity acquires a vacuum Berry curvature, an anomalous Hall conductivity, edge currents and an orbital magnetization~\cite{Tokatly2021VacuumAHE}, and in a quantum ring ultrastrong coupling to a gyrotropic cavity produces abrupt ground-state current jumps~\cite{Remolif2025}.  For a single charge, minimizing the photon energy at fixed $\mathbf p$ gives a displaced Gaussian photon state whose displacement is linear in $\mathbf p$, so the dressed band carries a uniform momentum-space Berry curvature,
\begin{equation}
\mathcal F_{xy}=\frac{4\Delta\,g^2\omega_0}{\big(\omega_+\omega_-+2g^2\omega_0\big)^2},\qquad \omega_\pm=\omega_0\pm\Delta,
\label{eq:berry}
\end{equation}
which is odd in the splitting $\Delta$, vanishes for a balanced
cavity, and is $O(g^2)$ at weak coupling \cite{Tokatly2021VacuumAHE} 
(see SM Sec.~S3.A).  A slowly applied force $\mathbf F$ then produces, besides the dressed 
drift, a transverse anomalous velocity $\mathcal F_{xy}|\mathbf F|$ on an exactly gapless band.  What the present hierarchy adds is the context: Eq.~\eqref{eq:nogo} shows that this Hall-like response coexists with commuting, gapless translations, and the same complete field, once reduced to one helicity (Case~III), loses the compensating Berry connection.  Two applications follow.  A transverse response with the sign and size of Eq.~\eqref{eq:berry} on a gapless ballistic band identifies the vacuum Berry-curvature mechanism of the complete dipole field, whereas a departure points beyond Case~II.  And the sign of $\mathcal F_{xy}$ is the sign of $\Delta$; together with the dressed-mass law (see SM Sec.~S3) it gives two parameter-free observables, $m^*(\Delta)$ and $\mathcal F_{xy}(\Delta)$, a self-consistency check on a cavity's coupling and splitting.

Case III: one retained helicity.  The third case deliberately changes the field content and compares two single-helicity reductions of the complete dipole Hamiltonian, which differ in whether the helicity projection precedes or follows the PZW transformation.  These reductions are not related by a gauge transformation and are not generally equivalent low-energy approximations; they are different effective models, not competing gauge predictions.  We now retain only one helicity $s=\pm1$ and represent its two in-plane components by the conjugate quadratures of one oscillator, as in Eq.~\eqref{eq:Acases}; we denote this potential $\hat{\mathbf A}_{1h}$, with $[A_x,A_y]=-isA_0^2$ and $qA_0=g$.
This is literally one helicity term of the complete field: in the basis \eqref{eq:circ}, $q\hat{\mathbf A}$ of Case~II, Eq.~\eqref{eq:Acases}, equals $\sum_{s=\pm}g(Q_s\hat{\mathbf x}-sP_s\hat{\mathbf y})$, and the noncommuting pieces of the two helicities cancel to give $[Q_x,Q_y]=0$ (see SM Sec.~S3).  Direct minimal coupling gives
\begin{equation}
{\hat H_{1h}^{V}=\frac12(p_x-gQ)^2+\frac12(p_y+sgP)^2+\frac{\omega_c}{2}(Q^2+P^2).}
\label{eq:oneV}
\end{equation}
This is the exact velocity-gauge Hamiltonian of the reduced one-helicity model, which we call Case III-V.  Despite $[A_x,A_y]\neq0$, it contains no $x$ or $y$, so $p_x$ and $p_y$ remain commuting constants of motion.  Writing $\Omega_c=\omega_c+g^2$, the electron drifts ballistically with isotropic mass $m_{1h}^*=\Omega_c/\omega_c$ plus bounded chiral micromotion. 

A special Landau-like theory, Case III-L, appears when the helicity projection is performed after, rather than before, the PZW transformation.  The complete unitary \eqref{eq:U2h} factorizes into single-helicity factors, $U=U_+U_-$ with $U_s=\exp[ig(xQ_s-syP_s)]$, and the complete length-gauge Hamiltonian is exactly a sum over helicities (see SM Sec.~S7.A); keeping one term of that sum, including its share of the dipole self-energy, gives
\begin{equation}
{\hat H_{1h}^{L,{\rm red}}=\frac{p_x^2+p_y^2}{2}+\frac{\omega_c}{2}\Big[(Q+sg\,y)^2+(P+g\,x)^2\Big],}
\label{eq:original3}
\end{equation}
equivalently $\tfrac12(P_c+s\lambda_t y)^2+\tfrac12(\omega_cQ_c-\lambda_t x)^2$ with $\lambda_t=\lambda/\sqrt2$ after a canonical relabeling.  Its structure is exactly that of a Landau problem coupled to a cavity oscillator.  The two nonzero normal-mode frequencies are
\begin{equation}
\omega_\mp=\frac{\sqrt{\omega_c^2+2\lambda^2}\mp\omega_c}{2},\qquad
\omega_+\omega_-=\frac{\lambda^2}{2},
\label{eq:freqs}
\end{equation}
and the remaining zero mode is a guiding centre.  The corresponding field scale is
\begin{equation}
{B_{\rm syn}=s g^2=s\frac{\lambda^2}{2\omega_c}.}
\label{eq:Bsyn}
\end{equation}
Hence the synthetic $B$, magnetic length, Landau-type degeneracy, bounded transverse dynamics, and Hall drift all survive exactly within Eq.~\eqref{eq:original3}.

The two models are related exactly (SM Secs.~S5 and S7.B).  Because $[Q,P]=i$, the Baker--Campbell--Hausdorff series of the PZW generator $g(xQ-syP)$ contains a residual term, and the exact PZW image of Eq.~\eqref{eq:oneV} is Eq.~\eqref{eq:original3} with $\mathbf p^2/2$ replaced by $(\mathbf p-q\mathbf A_B)^2/2$, $q\mathbf A_B=\tfrac12sg^2(y,-x)$.  Hence Eq.~\eqref{eq:original3} is the PZW image of the reduced velocity-gauge cavity supplemented by a real uniform field $qB_z=sg^2=B_{\rm syn}$ co-handed with the retained helicity; the counter-handed sign gives a different spectrum.  The Landau physics is therefore that of a single-helicity cavity plus a real field at the point $\omega_B=g^2$ of the Landau-polariton family $\hat H_{1h}^{V}+B$ (cf.\ the Landau polaritons of a two-dimensional electron gas in a cavity, known exactly in the optical limit~\cite{Rokaj2019QEDBloch,Rokaj2023WeakenedTopological}), or of a quadratic simulator that realizes the crossed couplings directly, not a property of the complete two-helicity vacuum.  The image of the ballistic model carries a residual field $-sg^2$ whereas the reduced Landau model carries $+sg^2$, so $H_{1h}^{V}$ and $H_{1h}^{L,{\rm red}}$ are different physical Hamiltonians, not alternative gauges of one.

\begin{figure}[t]
\centering
\includegraphics[width=\columnwidth]{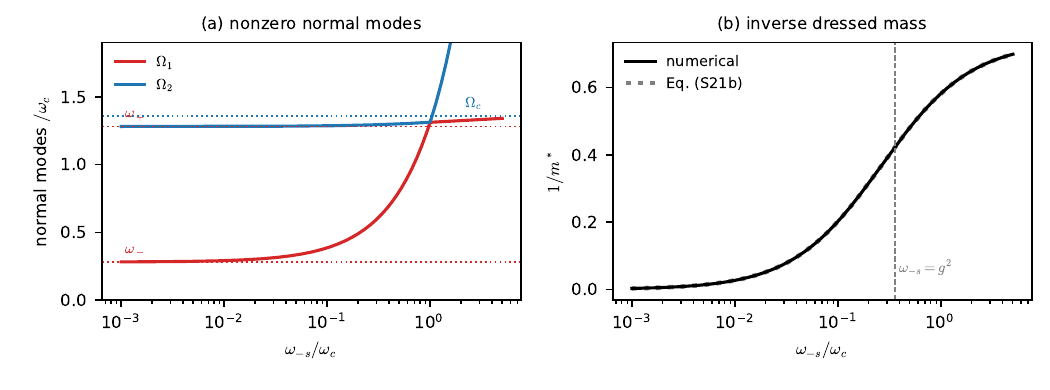}
\caption{Stiff-to-soft partner crossover of the complete gyrotropic theory ($g=0.6$, $\omega_s=\omega_c=1$). (a) Nonzero normal modes against the partner frequency $\omega_{-s}$: they approach $\omega_\mp$ of Eq.~\eqref{eq:freqs} (dotted red) as $\omega_{-s}\to0$, and the single-helicity value $\Omega_c=\omega_c+g^2$ (dotted blue) as $\omega_{-s}\to\infty$. (b) Inverse dressed mass $1/m^*$ (numerical, and the closed form of SM Sec.~S3): the translational band flattens as $\omega_{-s}/g^2\to0$, which is the Landau-like limit.}
\label{fig:crossover}
\end{figure}

Two further exact statements locate Eq.~\eqref{eq:original3} inside the complete theory (see SM Sec.~S7).  First, the complete length-gauge Hamiltonian reads
\begin{equation}
\hat H^{L}_{2h}=\frac{\mathbf p^2}{2}+\sum_{s=\pm}\frac{\omega_s}{2}\Big[(Q_s+sgy)^2+(P_s+gx)^2\Big],
\label{eq:sumrule}
\end{equation}
for any gyrotropic splitting $\omega_\pm$, with $[C_x,C_y]=-ig^2\sum_s s=0$: each helicity carries $B^{(s)}_{\rm syn}=sg^2$ and the two cancel.  Projection and gauge transformation do not commute, the field-side analogue of the gauge ambiguity of truncated matter spaces~\cite{DeBernardis2018,DiStefano2019,Stokes2019,Taylor2020}.  Second, the fate of $B_{\rm syn}$ is decided by the frequency of the partner helicity retained in the complete theory.  If the partner is stiff, $\omega_{-s}\gg g^2$, its displaced vacuum follows the electron adiabatically and contributes the Berry connection $q\mathbf A_B$ that is the residual term, so the effective theory is the ballistic model, Eq.~\eqref{eq:oneV}.  If the partner is softened, $\omega_{-s}\to0$, its quadratures enter only through $\bm\pi=(p_x-gQ_{-s},\,p_y-sgP_{-s})$ with $[\pi_x,\pi_y]=isg^2$ and supply the algebra of a real uniform field $qB_z=sg^2$.  An explicit invertible $8\times8$ symplectic map (SM Sec.~S7.D) then decomposes the soft-point Hamiltonian into the same two-pair active Landau-polariton Hamiltonian as Eq.~\eqref{eq:original3} together with two decoupled spectator canonical pairs.  The equivalence is therefore one of active sectors, not of physical observables: the electron position depends on the spectators, and the electron--photon entropy of the complete theory ($3.02$ nats against $2.88$ for $g=0.6$, $\omega_s=1$, $\omega_st\le60$) depends on the partner preparation (see SM Sec.~S10.B).  For small finite $\omega_{-s}$ the translations still commute but $m^*\to\infty$, the band flattening as $E(\mathbf p)\simeq\omega_{-s}\mathbf p^2/(2g^2)$, so the Landau ladder is recovered as the flat-band limit, with the two nonzero normal modes converging to $\omega_\mp$ of Eq.~\eqref{eq:freqs} (Fig.~\ref{fig:crossover} and SM Table~II).

The controlling parameter is thus the ratio $\omega_{-s}/g^2$, not the spectral purity with which the unwanted helicity is suppressed.  The stiff-partner limit is the physical justification for the ballistic model and, for a helicity-selective cavity that pushes the unwanted helicity out of band, the expected effective theory (III-V).  The Landau-like active sector instead emerges through an exact canonical reorganization at the singular soft-partner point.  Projecting after the PZW transformation is not by itself a physical mechanism for synthetic magnetism: the Landau-like form (III-L) otherwise requires a real co-handed field or engineered crossed couplings, and its realization in a microscopic chiral cavity remains to be demonstrated.

This hierarchy also clarifies what is meant by a single-handed cavity.  Optical structures with an eigenmode of one handedness in a selected spectral window~\cite{Voronin2022SingleHandedness}, twisted conducting resonators with helical TE--TM hybrid modes~\cite{Paterson2025HelicityTwisted}, and high-helicity superconducting resonators~\cite{Crew2026UPLOADHELIX} make the field content of a ``chiral cavity'' a physical, rather than merely notational, question.  They motivate a single-helicity low-energy description but do not by themselves prove that the complete canonical field may be replaced by the Case~III potential before a gauge transformation.  Equation~\eqref{eq:original3} should therefore be regarded as a distinguished effective limit: a single-helicity cavity with a real co-handed uniform field $qB_z=s\lambda^2/(2\omega_c)$, or the active sector of the singular soft-partner limit of a complete gyrotropic theory.  A helicity-selective resonator that pushes the partner out of band realizes instead the ballistic model, Eq.~\eqref{eq:oneV}.  Active modulation of a non-magnetic cavity anisotropy relocates rather than removes the real-field-equivalent physics (SM Sec.~S7.D.1), and the one genuinely field-free route engineers the crossed couplings directly, for example via a gyrator or a gyroscopic coupling (SM Table~IV).

{\it Significance and applications} (Case III).  Case III locates exactly where a single-helicity description of a chiral cavity is and is not justified, which makes the partner-helicity frequency the controlling experimental knob.  Tuning it through $\omega_{-s}/g^2\sim1$ is a mode-content experiment that switches a charge between ballistic and orbital dynamics.  Equation~\eqref{eq:original3} is also an analytically exact quadratic simulator for guiding-centre algebra, cyclotron motion and Hall drift, one point $qB=sg^2$ of the solvable family $\hat H_{1h}^V+B$, whose field sweep gives an analytic calibration curve for $g$ and $\omega_c$ (see SM Fig.~2).  Dissipative extensions are model dependent and are discussed in the SM.

Every Hamiltonian above is quadratic in the electron and photon canonical variables, so one symplectic propagator solves all of them in closed form: Heisenberg equations, spectra, Gaussian wavepacket dynamics and covariances, without numerical time propagation (SM Sec.~S10).  In the worked example ($g/\omega_c=0.6$, SM Fig.~4), Cases I, II and III-V show ballistic drift with bounded micromotion, whereas the reduced Hamiltonian~\eqref{eq:original3} gives a bounded cyclotron-like orbit with bounded $\mathrm{Var}\,x$, and the electron--photon entropy is bounded and oscillatory in all four models, largest for the Landau-like reduction.  These trajectories and spectra are direct experimental discriminants between complete-field and single-helicity descriptions; the entropy depends on the chosen subsystem variables, whereas physical trajectories are gauge invariant (SM Secs.~S10 and S11).

The three cases give a unified analytic picture.  Linear polarization selects one matter axis.  A complete transverse field, in Cartesian or helicity variables and with or without helicity splitting, preserves commuting translations and cannot magnetize a free charge in the dipole limit.  A single-helicity reduction has two inequivalent forms, obtained by reducing in velocity gauge or after the PZW transformation: the first is gapless and ballistic, whereas the second, Eq.~\eqref{eq:original3}, supports the synthetic field~\eqref{eq:Bsyn} and the associated Landau phenomenology.  They are related exactly by a real co-handed uniform field, the complete gyrotropic theory connects them through the singular soft-partner limit, and the stiff-partner limit, the one expected for a helicity-selective cavity, restores the ballistic dynamics.  Mode-content tuning, controlled partner softening and finite-$k$ structured fields are concrete routes to test which description is realized.  All results are for the dipole limit, which requires the packet to be small compared with the cavity wavelength; at finite wavevector the orbital magnetic-dipole term $-(q/2m)\mathbf L\cdot\mathbf B$ enters, and the no-go result should not be extrapolated beyond it (see SM Sec.~S13).  Derivations and further results are given in the Supplemental Material.

\renewcommand{\acknowledgmentsname}{ACKNOWLEDGMENT}
\begin{acknowledgments}
This work was supported by the National Science Foundation (NSF) under Grant No.\ DMR-2217759.
\end{acknowledgments}

\end{document}